\documentclass[fleqn]{article}
\usepackage{fleqn}
\usepackage{graphicx}
\usepackage{epsfig}
\usepackage{amsmath}
\usepackage{amssymb}
\usepackage{calc}
\newcommand{\beq}{\begin{equation}}
\newcommand{\eeq}{\end{equation}}
\newcommand{\bea}{\begin{eqnarray}}
\newcommand{\eea}{\end{eqnarray}}
\begin{document}
\Large
\begin{center}
{\bf  Three-Qubit Operators, the Split Cayley Hexagon of Order Two and Black Holes}
\end{center}
\large
\vspace*{-.1cm}
\begin{center}
P\'eter L\'evay,$^{1}$ Metod Saniga$^{2}$ and P\'eter Vrana$^{1}$

\end{center}
\vspace*{-.4cm} \normalsize
\begin{center}
$^{1}$Department of Theoretical Physics, Institute of Physics, Budapest University of\\ Technology and Economics, H-1521 Budapest, Hungary

\vspace*{.0cm} and

\vspace*{.0cm}

$^{2}$Astronomical Institute, Slovak Academy of Sciences\\
SK-05960 Tatransk\' a Lomnica, Slovak Republic

\vspace*{.3cm} (11 September 2008)

\end{center}

\vspace*{-.3cm} \noindent \hrulefill

\vspace*{.1cm} \noindent {\bf Abstract}

\noindent The set of 63 real generalized Pauli matrices of
three-qubits can be factored into two subsets of 35 symmetric and
28 antisymmetric elements. This splitting is shown to be
completely embodied in the properties of the Fano plane; the
elements of the former set being in a bijective correspondence
with the 7 points, 7 lines and 21 flags, whereas those of the
latter set having their counterparts in 28 anti-flags of the
plane. This representation  naturally extends to the one in terms
of the split Cayley hexagon of order two. 63 points of the hexagon
split into 9 orbits of 7 points (operators) each under the action
of an automorphism of order 7. 63 lines of the hexagon carry three
points each and represent the triples of operators such that the
product of any two gives, up to a sign, the third one. Since this
hexagon admits a full embedding in a projective 5-space over
$GF(2)$, the 35 symmetric operators are also found to answer to
the points of a Klein quadric in such space. The 28 antisymmetric
matrices can be associated with the 28 vertices of the Coxeter
graph, one of two distinguished subgraphs of the hexagon. The
$PSL_{2}(7)$ subgroup of the automorphism group of the hexagon is
discussed in detail and the Coxeter sub-geometry is found to be
intricately related to the $E_7$-symmetric black-hole entropy
formula in string theory. It is also conjectured that the full
geometry/symmetry of the hexagon should manifest itself in the
corresponding black-hole solutions. Finally, an intriguing analogy
with the case of Hopf sphere fibrations and a link with coding
theory are briefly mentioned.

\vspace*{.3cm}
\noindent
{\bf PACS:} 02.40.Dr, 03.65.Ud, 03.65.Ta, 03.67.-a, 04.70.-s\\
{\bf Keywords:}  Three-Qubit Pauli Group --- Fano Plane  --- Generalized Hexagon of Order Two\\ \hspace*{1.95cm} --- $E_7$-Symmetric Stringy Black Holes

\vspace*{-.2cm} \noindent \hrulefill

\section{Introduction}
Entanglement and geometry are key concepts in contemporary
theoretical physics. The first can be regarded as the
``characteristic trait of quantum theory", the second as the basic
unifying agent in explaining the fundamental forces. A vital
combination of these concepts shows up in two seemingly unrelated
fields, quantum information and quantum gravity. The basic goal in
quantum information is understanding entanglement in geometric
terms, and in quantum gravity the understanding of the structure
of the Bekenstein-Hawking entropy of certain stringy black hole
solutions. Recently, in a series of papers a striking
correspondence has been established between this two apparently
separate fields \cite{Duff1}--\cite{Duff3}. For the intriguing
mathematical coincidences underlying this correspondence M. J.
Duff and S. Ferrara coined the term ``Black Hole Analogy". The
basic correspondence of the analogy is the one between the black
hole entropy formula of certain stringy black hole solutions on
one hand and entanglement measures for qubit and qutrit systems on
the other. The archetypical example \cite{Duff1} of such a
correspondence is the one concerning  the $8$-charge STU black
hole of $N=2$, $D=4$ supergravity \cite{Liu,Behrndt}. The entropy
formula in this case is of the form \beq
S=\frac{\pi}{2}\sqrt{\tau_{ABC}}, \label{stu} \eeq \noindent where
${\tau}_{ABC}$ is the so called three-tangle \cite{Kundu}, an
entanglement measure characterizing entangled  three-qubit systems
$A,B$ and $C$. Later works \cite{Linde,Ferrara,Levay4} have
revealed that the most general type of stringy black hole
solutions with $56$ charges occurring within the context of $N=8$,
$D=4$ supergravity also display this correspondence. Here the
$E_{7(7)}$ symmetric black hole entropy formula is \beq
S=\frac{\pi}{2}\sqrt{\tau_{ABCDEFG}}, \label{N8} \eeq \noindent
where $\tau_{ABCDEFG}$ is an entanglement measure characterizing
the {\it tripartite entanglement of seven qubits}. This measure is
related to the unique quartic Cartan invariant for the fundamental
representation of the exceptional group $E_{7(7)}$, and to the
discrete geometry of the Fano plane. Truncations of the discrete
$U$-$duality$ group $E_{7}({\bf Z})$ have been connected to
truncations of the Fano plane to its points, lines and quadrangles
\cite{Levay4}. This result can be regarded as the first indication
that discrete geometric ideas could play a key role in a
quantum-entanglement-based understanding of duality symmetries. As
a latest development, a physical basis for the Black Hole Analogy
in the special case of the STU model has recently been revealed.
It has been shown that the qubits occurring in the analogy can be
realized as wrapped intersecting D3-branes \cite{Duff3}.

It has become obvious from the very beginning that the quantum
information theoretic basis for the black hole  correspondence
rests on the very special geometric properties of few qubit
systems \cite{Levay1}. Indeed, in the quantum entanglement
literature it has become common wisdom that the special geometry
of one, two and three-qubit systems can naturally be related to
Hopf fibrations related to the complex numbers, quaternions and
octonions, respectively \cite{Mosseri}--\cite{Bernevig}. These
approaches to few qubit systems studied the geometry of
entanglement via looking at the geometry of the corresponding
multi-qubit {\it Hilbert space}. An alternative approach to
understanding entanglement in geometric terms is provided by the
one of studying the geometry of the {\it algebra of observables}
with special operators (e.\,g., CNOT operations) creating and
destroying entanglement by acting on it. In particular, it proved
to be of fundamental importance to look at the structure of the
so-called Pauli group on few qubits. This group is of utmost
importance in the field of fault tolerant quantum computation and
stabilizer codes. The algebra of the generalized Pauli group of
$N$-qubit systems has also remarkable discrete geometric
properties, being completely embodied in the so-called {\it
symplectic polar spaces} of order two and rank $N$
\cite{Metod1}--\cite{hav}.

Although the appearance of few qubit (and qutrit) entangled
systems related to discrete geometric structures provided an
additional insight into the structure of stringy black hole
solutions, some basic points still remained obscure. In
particular, the entangled state characterizing the tripartite
entanglement of seven qubits is not the one occurring in
conventional quantum information theory. This state is {\it not}
an element of any subspace of the Hilbert space on seven qubits.
Indeed, it has been shown that it is an element of a
$56$-dimensional subspace of seven qutrits \cite{Ferrara}.
Moreover, a special role of three-qubit systems repeatedly showing
up in these systems hints at an alternative structure lurking
behind the scene.

Motivated by the desire to eliminate such shortcomings, in this
paper we initiate an observable-based understanding of the Black
Hole Analogy. Our basic object of scrutiny is the discrete
geometry of the Pauli group of three-qubits. Although the
corresponding geometry of two-qubits, which is that of the {\it
generalized quadrangle of order two}  \cite{Metod2,plasan}, has
already been worked out in detail, a similar comprehensive
analysis for three-qubits is still missing. The main body of the
paper is, therefore, devoted to a detailed study of the geometry
of the {\it real} operators of this group. The principal result of
this part is an explicit demonstration of the fact that the $63$
real operators of the Pauli group can be mapped bijectively to the
points of the so-called {\it split Cayley hexagon of order two}.
The last part of the paper is devoted to establishing an explicit
correspondence between a sub-geometry of this hexagon and the
$E_{7(7)}$ symmetric black hole entropy formula. The symmetry of
this (Coxeter-graph-based) geometry turns out to be related to a
$PSL_{2}(7)$ subgroup of the full $U$-duality group. In order to
relate our observable based approach to the seven-qubit Hilbert
space based one \cite{Ferrara,Levay4}, we provide the basic
dictionary relating these two approaches. Finally, we also
conjecture that the geometry and symmetry of the Hexagon in its
full generality should manifest itself at the level of such black
hole solutions. Our hope is that this new approach will initiate
further investigations to fill in the missing gaps, and to explore
the possible physical consequences.

The paper is organized as follows. First, we shall demonstrate
that the algebra of our 63 generalized Pauli matrices is fully
encoded in the properties of the Fano plane and its dual
(Sec.\,2). Then, by employing its automorphism of order 7
(Sec.\,3), we shall illustrate in many details how this encoding
naturally and straightforwardly translates into the properties and
characters of the points/lines of this split Cayley hexagon and
one of its representation inside a projective 5-space (Sec.\,4).
Next, a quite in-depth examination of $PSL_{2}(7)$, a subgroup of
the full automorphism group of the hexagon, will follow (Sec.\,5).
Finally, the Coxeter graph, one of two most important
sub-configurations living within our hexagon (the other being the
incidence graph of the Fano plane --- the Heawood graph), will be
given close scrutiny and shown to be intricately linked with the
$E_7$-symmetric black-hole entropy formula in string theory
(Sec.\,6). The paper will be finished by a brief discussion of an
intriguing analogy with Hopf sphere fibrations and an outline of a
worth-exploring link with coding theory.

\section{Real operators on three-qubits and the Fano plane}
Let us define the following set of $2\times 2$ matrices
\begin{equation}
I=\begin{pmatrix} 1&0\\0&1\end{pmatrix},\quad X=\begin{pmatrix} 0&1\\1&0\end{pmatrix},\quad Y=\begin{pmatrix}0&1\\-1&0\end{pmatrix},\quad
Z=\begin{pmatrix}1&0\\0&-1\end{pmatrix}.
\label{paulimatrices}
\end{equation}
\noindent
Notice that except for $Y$ these are the $2\times 2$ identity and the usual
Pauli spin matrices.
They are symmetric except for $Y$ which is antisymmetric.
We note that the set ${\cal M}=\{\pm I, \pm X,\pm Y,\pm Z\}$ consists of the {\it real} operators of the Pauli group acting on a single qubit.

We consider three-qubit systems with Hilbert space ${\cal H}\equiv
{\bf C}^2\otimes {\bf C}^2\otimes {\bf C}^2$ on which  the
Kronecker (tensor) product of three $2\times 2$ matrices $A\otimes B\otimes
C$ is defined in the usual way. Here we refer to $A$, $B$ and $C$
as the matrices acting on the first, second and third qubit,
respectively. Let us now assume that $A,B,C \in {\cal M}$ and
introduce the shorthand notation $ABC$ for their Kronecker product;
hence, for example, we have

\beq
ZYX\equiv Z\otimes Y\otimes X=\begin{pmatrix}Y\otimes X&0\\0&-Y\otimes X\end{pmatrix}=\begin{pmatrix}0&X&0&0\\-X&0&0&0\\0&0&0&-X\\0&0&X&0\end{pmatrix}.
\label{example}
\eeq
\noindent
Notice that the $8\times 8$ matrices containing an {\it odd} number of $Y$'s
are {\it antisymmetric}, whilst those comprising an {\it even} number of $Y$'s are {\it symmetric}; thus, for example, $ZYX$ is an antisymmetric matrix, but  $IYY$
and $XZX$ are symmetric ones.

Up to a sign, we have altogether $4^3=64$ distinct $8\times 8$
matrices. After omitting the identity $III\equiv I\otimes I\otimes
I$ we are left with $63$ matrices of which $28$ are antisymmetric
and $35$ symmetric. In the space of symmetric $8\times 8$
matrices we consider the following two seven-element sets

\beq {\cal L} \equiv\{IIX,IXI,IXX,XII,XIX,XXI,XXX\} \label{point}
\eeq
\noindent
and
\beq {\cal P}
\equiv\{IIZ,IZI,IZZ,ZII,ZIZ,ZZI,ZZZ\} \label{line} \eeq
\noindent
to have a special status. We will see soon that the elements of these sets
can be associated with the {\it points} (${\cal P})$ and {\it lines}
(${\cal L}$) of the projective plane of order two, the Fano plane. Here we simply observe that the elements of these sets are pairwise commuting.

Our next step is to create a multiplication table by taking the
elements of the set ${\cal L}$ to label the rows and those of the
set ${\cal P}$ to label the columns of a $7\times 7$ array whose
entries are, {\it up to a sign}, the (ordinary matrix) products of
the corresponding elements. After carrying out all the
calculations, we notice that the 21 symmetric and the 28
antisymmetric matrices from this $7\times 7$ array reveal an
incidence structure reminiscent of the one of the Fano plane. More
precisely, we find three pairwise commuting symmetric and four
pairwise commuting antisymmetric combinations occurring in both
each row and each column. Moreover, any two elements such that one
belongs to the symmetric and the other to the antisymmetric set
(lying in a particular row or column) are anti-commuting (not
commuting). Next, if we look at the pattern of the location of
symmetric combinations pertaining to different rows and columns we
obtain the same structure we get by looking at the incidence
structure of points and lines of the Fano plane and its dual.

\begin{table}[t]
\begin{center}
\caption{The matrix products between the operators from the two distinguished sets.}
\vspace*{0.3cm}
\begin{tabular}{||l|lllllll||}
\hline \hline
&&&&&&&\\[-.2cm]
Lines/Points & $ZZI$ & $ZII$ & $ZZZ$ & $IZI$ & $IZZ$ & $ZIZ$ & $IIZ$ \\[1.mm]
\hline
&&&&&&&\\[-.2cm]
$IIX$ & $ZZX$ & $ZIX$ & \underline{$ZZY$} & $IZX$ & \underline{$IZY$} & \underline{$ZIY$} & \underline{$IIY$} \\
$IXX$ & \underline{$ZYX$} & $ZXX$ & $ZYY$ & \underline{$IYX$}& $IYY$ & \underline{$ZXY$} & \underline{$IXY$} \\
$XIX$ & \underline{$YZX$} & \underline{$YIX$} & $YZY$ & $XZX$ & \underline{$XZY$} & $YIY$ & \underline{$XIY$} \\
$XII$ & \underline{$YZI$} & \underline{$YII$} & \underline{$YZZ$} & $XZI$ & $XZZ$ & \underline{$YIZ$} & $XIZ$ \\
$XXX$ & $YYX$ & \underline{$YXX$} & \underline{$YYY$} & \underline{$XYX$} & $XYY$ & $YXY$ & \underline{$XXY$} \\
$IXI$ & \underline{$ZYI$} & $ZXI$ & \underline{$ZYZ$} & \underline{$IYI$} & \underline{$IYZ$} & $ZXZ$ & $IXZ$ \\
$XXI$ & $YYI$ & \underline{$YXI$} & $YYZ$ & \underline{$XYI$} & \underline{$XYZ$} & \underline{$YXZ$} & $XXZ$ \\[1.mm]
\hline \hline
\end{tabular}
\end{center}
\end{table}

Let us explicitly see how this works. After a particular ordering
of the elements of the sets ${\cal P}$ and ${\cal L}$ we can group
the remaining 49 matrices in the manner as depicted in Table 1.
In the first row ($IIX$) we have pairwise
commuting symmetric combinations in columns  $1, 2,$ and $4$
(i.\,e., in the columns labelled by the triples $ZZI,ZII,$ and
$IZI$), and pairwise commuting antisymmetric ones in columns $3,
5,6,$ and $7$ (that is, in $ZZZ, IZZ, ZIZ,$ and $IIZ$ ---
the corresponding entries being underlined). Looking at the other rows we
readily notice a {\it cyclic shift} of this pattern: $(124)$
$\rightarrow$ $(235)$ $\rightarrow$ $(346)$ $\rightarrow$ $(457)$
$\rightarrow$ $(561)$ $\rightarrow$ $(672)$ $\rightarrow$ $(713)$.
It represents no difficulty to check that this pattern is embodied
in the structure of the Fano plane, as shown in Fig. 1, left;
here, the grey numbers (points) respectively black ones (lines)
from 1 to 7 label the columns respectively rows, together with the
respective matrices of the two special sets, of Table 1. Dually,
in the first column ($ZZI$) we have pairwise commuting symmetric
combinations in rows $5, 7,$ and $1$ ($XXX$, $XXI$, $IIX$),
and pairwise commuting antisymmetric ones in rows $2, 3, 4,$ and
$6$ ($IXX$, $XIX$, $XII$, and $IXI$). Now the cyclically shifted
pattern looks as follows $(134)$ $\rightarrow$ $(245)$
$\rightarrow$ $(356)$ $\rightarrow$ $(467)$ $\rightarrow$ $(571)$
$\rightarrow$ $(612)$ $\rightarrow$ $(723)$, being reproduced by
the geometry of the dual Fano plane (Fig.\,1, right).
\begin{figure}[pth!]
\centerline{\includegraphics[width=14.0truecm,clip=]{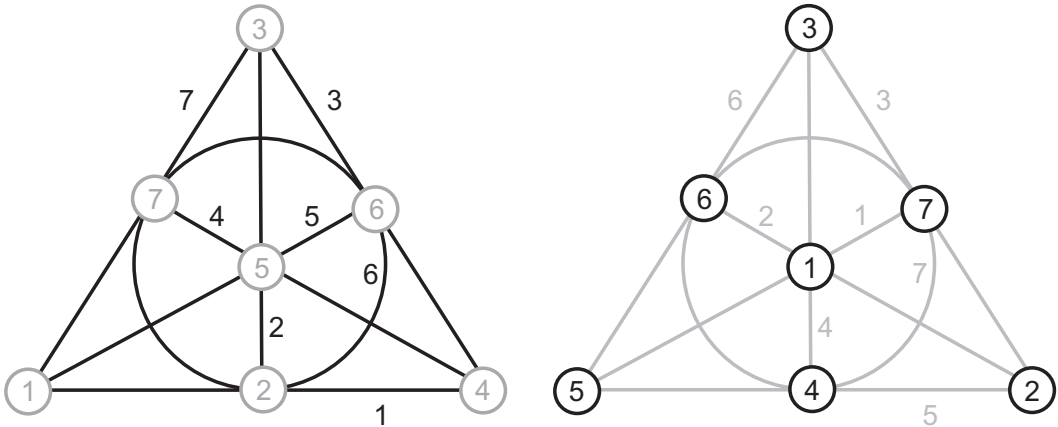}}
\caption{The Fano plane and its dual as the fundamental building blocks of the three-qubit
Pauli group algebra.}
\end{figure}

\begin{figure}[pth!]
\centerline{\includegraphics[width=10truecm,clip=]{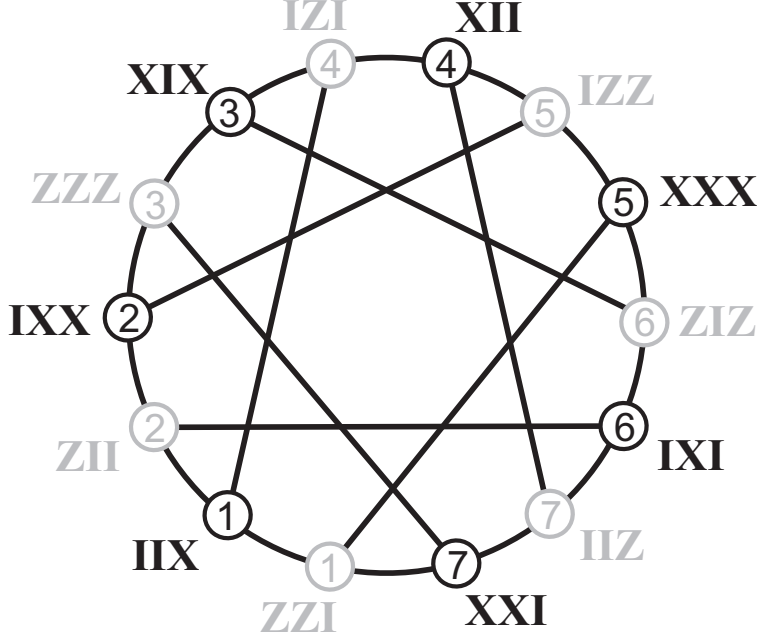}}
\caption{The algebra of the operators from the two distinguished sets in terms
of the incidence graph of the Fano plane.}
\end{figure}

This dual view of the operators' algebra can be cast into a single compact form
by employing the incidence graph of the Fano plane, also known as the Heawood graph
(Fig.\,2). This graph consists of  $14$ vertices corresponding to the seven points (grey circles)
and seven lines (black circles) of the Fano plane, with two vertices (of different shading) being
adjacent if they correspond to a point and a line such that the point is contained in the line.

From the above-outlined picture we observe that the symmetric
combinations correspond to the flags (a line taken together with a
point lying on it) and the antisymmetric ones to the antiflags (a
line and a point {\it not} lying on that line) of the Fano plane.
Taken together with the sets ${\cal P}$ and ${\cal L}$ whose
elements label the points and the lines, we see that {\it all}
$63$ nontrivial matrices forming a subset of the generalized
three-qubit Pauli group can {\it completely} be described in
geometrical terms related to the structure of the Fano plane. This
is a very important finding because there exists a remarkable
unique finite geometry whose properties can be fully derived from
the structure of the Fano plane {\it alone}, viz. a thick
generalized polygon called split Cayley hexagon
\cite{Polster1}--\cite{Maldeghem}. In order to show that this
geometry indeed mimics all the essential features of the
commutation algebra of our generalized Pauli matrices, one more
piece of three-qubit mathematics must be given a careful inspection and
properly understood --- automorphisms of order 7.

\section{Automorphism of order seven and its orbits}

To this end in view, let us denote by $\alpha$ the following
permutation: $(1234567)$; this is obviously of order seven. We
would like to find an $8\times 8$ representation of $\alpha$ which
permutes cyclically the elements of the ordered sets ${\cal P}$
and ${\cal L}$ we used to label the rows and columns of Table 1.
We have to find, in particular, an $8\times 8$ matrix ${\cal
D}(\alpha)$ with the property $({\cal D}(\alpha))^7=III$ that acts
on the elements of ${\cal P}$ and ${\cal L}$ by conjugation in the
following way \beq {\cal D}(\alpha): IIX\mapsto IXX\mapsto
XIX\mapsto XII\mapsto XXX\mapsto IXI\mapsto XXI\mapsto
IIX\mapsto\dots, \label{laction} \eeq \beq {\cal D}(\alpha):
ZZI\mapsto ZII\mapsto ZZZ\mapsto IZI\mapsto IZZ\mapsto ZIZ\mapsto
IIZ\mapsto ZZI\mapsto\dots, \label{paction} \eeq \noindent where
\beq {\cal D}(\alpha)^{-1}IIX{\cal D}(\alpha)=IXX, \quad {\cal
D}(\alpha)^{-1}IXX{\cal D}(\alpha)=XIX,\quad\dots. \label{conj}
\eeq Such a matrix is of the form \beq {\cal
D}(\alpha)\equiv\begin{pmatrix}P&Q&0&0\\0&0&Q&P\\0&0&QX&PX\\PX&QX&0&0\end{pmatrix},\quad
P=\begin{pmatrix}1&0\\0&0\end{pmatrix},\quad
Q=\begin{pmatrix}0&0\\0&1\end{pmatrix}. \label{dalfa} \eeq
\noindent Now let us look at the action of ${\cal D}(\alpha)$ on
$21$ symmetric matrices labelling the flags and 28 antisymmetric
matrices labelling the anti-flags of the Fano plane. We find
exactly seven orbits of seven elements each. The aggregate of $21$
symmetric matrices splits into three orbits. One of them is
represented by the elements located on the main diagonal of Table
1, \beq {\cal D}(\alpha): ZZX\mapsto ZXX\mapsto YZY\mapsto
XZI\mapsto XYY\mapsto ZXZ\mapsto XXZ\mapsto ZZX\mapsto\dots,
\label{main} \eeq \noindent the other two being represented by the
matrices sitting at the positions shifted rightward by one column
and/or three columns from the main diagonal in a cyclic manner,
namely \beq {\cal D}(\alpha): ZIX\mapsto ZYY\mapsto XZX\mapsto
XZZ\mapsto YXY\mapsto IXZ\mapsto YYI\mapsto ZIX\mapsto\dots,
\label{shift1} \eeq and \beq {\cal D}(\alpha): IZX\mapsto
IYY\mapsto YIY\mapsto XIZ\mapsto YYX\mapsto ZXI\mapsto YYZ\mapsto
IZX\mapsto\dots, \label{shift3} \eeq respectively. The set of $28$
antisymmetric matrices factors into four orbits; the elements of
all of them are located above the main diagonal shifted from it by
two, four, five and six columns. Their explicit forms read \beq
{\cal D}(\alpha): ZZY\mapsto IYX\mapsto XZY\mapsto YIZ\mapsto
XXY\mapsto ZYI\mapsto YXI\mapsto ZZY\mapsto\dots, \label{shift2}
\eeq \beq {\cal D}(\alpha): IZY\mapsto ZXY\mapsto XIY\mapsto
YZI\mapsto YXX\mapsto ZYZ\mapsto XYI\mapsto IZY\mapsto\dots,
\label{shift4} \eeq \beq {\cal D}(\alpha): ZIY\mapsto IXY\mapsto
YZX\mapsto YII\mapsto YYY\mapsto IYI\mapsto XYZ\mapsto
ZIY\mapsto\dots, \label{shift5} \eeq and \beq {\cal D}(\alpha):
IIY\mapsto ZYX\mapsto YIX\mapsto YZZ\mapsto XYX\mapsto IYZ\mapsto
YXZ\mapsto IIY\mapsto\dots, \label{shift6} \eeq respectively.

It is important to realize here that the action of ${\cal D}(\alpha)$ on the
three-qubit Hilbert space changes, in general, the entanglement
type (see, e.\,g., the SLOCC classes a particular state belongs
to \cite{Dur}). This can be demonstrated by expressing ${\cal
D}(\alpha)$ in terms of three-qubit CNOT operations, which are
well known to create or destroy entanglement \cite{Nielsen}. One possible way to
express ${\cal D}(\alpha)$ in terms of CNOT operations is \beq
{\cal D}(\alpha)=(C_{12}C_{21})(C_{12}C_{31})C_{23}(C_{12}C_{31}).
\label{fifth} \eeq \noindent Here, the explicit forms of the
three-qubit CNOT operations are \cite{Nielsen}
\beq
C_{12}=\begin{pmatrix}I&0&0&0\\0&I&0&0\\0&0&0&I\\0&0&I&0\end{pmatrix},\qquad
C_{21}=\begin{pmatrix}I&0&0&0\\0&0&0&I\\0&0&I&0\\0&I&0&0\end{pmatrix},
\label{cnots1} \eeq \noindent \beq
C_{23}=\begin{pmatrix}I&0&0&0\\0&X&0&0\\0&0&I&0\\0&0&0&X\end{pmatrix},
\qquad
C_{31}=\begin{pmatrix}P&0&Q&0\\0&P&0&Q\\Q&0&P&0\\0&Q&0&P\end{pmatrix},
\label{cnots2} \eeq \noindent
with the first index relating to
the location of the control bit and the second one to the target.

The fourth orbit of antisymmetric matrices, Eq.\,(\ref{shift6}),
has several noteworthy properties. The seven operators of this
orbit are pairwise anti-commuting. Since the square of each of
these operators is $-III$, replacing $Y$ by $\sigma_2=-iY$ we get
the matrices forming a representation of a seven dimensional
Clifford algebra, \beq \{i{\Gamma}_1,i{\Gamma}_2,
i{\Gamma}_3,i{\Gamma}_4,i{\Gamma}_5,i{\Gamma}_6,i{\Gamma}_7\}=\{IIY,ZYX,YIX,YZZ,XYX,IYZ,YXZ
\} \eeq \beq {\Gamma}_j{\Gamma}_k + {\Gamma}_k{\Gamma}_j =
2{\delta}_{jk}{\bf 1},\qquad {\bf 1}\equiv III,\qquad
j,k=1,2,\dots, 7. \label{clifford} \eeq It is then straightforward
to check that the remaining three orbits of antisymmetric matrices
can be generated as the commutators of the form
$\frac{1}{2}[{\Gamma}_j,{\Gamma}_k]$. This shows that the
remaining $21$ matrices, Eqs.\,(\ref{shift2})--(\ref{shift5}),
form the irreducible spinor representation of an $so(7)$ algebra.
Hence, the totality of $28$ antisymmetric matrices
$S_{0k}=\frac{i}{2}{\Gamma}_k$ and
$S_{jk}=\frac{1}{4}[{\Gamma}_j,{\Gamma}_k]$, associated with the
anti-flags of the Fano plane, form the generators of one of the
irreducible spinor representations of the $so(8)$ algebra with the
generators $S_{jk}$ giving rise to the $so(7)$ subalgebra.

Finally, we will notice that the elements of our specific fourth orbit can
be used to label the points of an {\it oriented} dual Fano plane.
That is, instead of the ordered set ${\cal
L}=\{IIX,IXX,XIX,XII,XXX,IXI,XXI\}$ labelling the rows of Table 1
and the points of Fig. 1, right, we can use the ordered set ${\cal
L}^{\prime}=\{IIY,ZYX,YIX,YZZ,XYX,IYZ,YXZ\}$. Since, for example,
the result of the multiplication $(XYX)(YZZ)(ZYX)$ (reading this
from left to right) gives $+IZZ$, we can endow the line
labelled by $IZZ$ with an arrow pointing from left to
right. Due to the above-described Clifford property, all even
permutations of this triple product result in $+IZZ$, while odd
ones yield $-IZZ$. Proceeding this way, every line of the dual
Fano plane can be given a unique orientation --- as illustrated in Fig.\,3.
Making one step further and employing the following correspondence between
the unit octonions and the elements of our special set,
$\{e_1,e_2,e_3,e_4,e_5,e_6,e_7\} \leftrightarrow
\{XYX,YZZ,YIX,ZYX,IIY,YXZ,IYZ\}$, Fig.\,3  is found to be identical with a simple mnemonic
for the products of the unit octonions \cite{Baez}. Note, however, that for the octonions
we have, for example, $e_1e_2=e_4$, but in our case $(XYX)(YZZ)\neq ZYX$. So,
the set ${\cal L}^{\prime}$ can only  be used to mimic the
signs occurring in the octonionic multiplication table.
\begin{figure}[pth!]
\centerline{\includegraphics[width=10truecm,clip=]{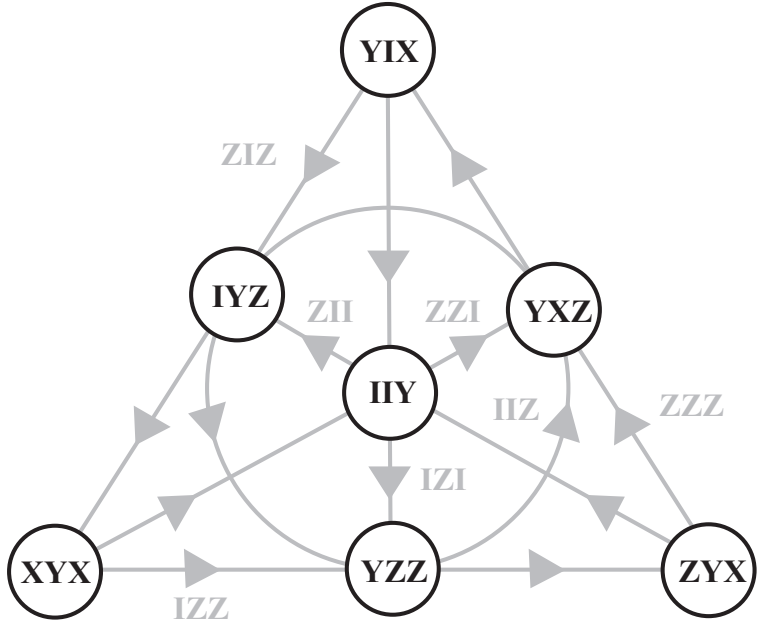}}
\caption{The oriented dual Fano plane.}
\end{figure}

At this point we have at hand all the necessary technicalities to reveal the smallest split Cayley hexagon
lurking behind the core algebraic
properties of the $63$ nontrivial real operators of the Pauli
group on three qubits.

\section{Core geometry: the smallest split Cayley hexagon}

A finite generalized $n$-gon ${\cal G}$ of
order $(s,t)$ is a point-line incidence geometry satisfying the following
three axioms \cite{Polster1}--\cite{Maldeghem}:

(A1) Every line contains $s+1$ points and every point is contained in $t+1$ lines.

(A2) ${\cal G}$ does not contain any ordinary $k$-gons for $2\leq k<n$.

(A3) Given two points, two lines, or a point and a line, there is at least one ordinary

$n$-gon in ${\cal G}$ that contains both objects.

\noindent From the definition it is obvious that the dual of a
generalized $n$-gon is also a generalized $n$-gon. If $s=t$ we say
that the $n$-gon is of order $s$. An ordinary $n$-gon is the
unique generalized $n$-gon of order $1$. A generalized polygon is
called {\it thick} if every point is contained in at least three
lines and every line contains at least three points, i.\,e., if
$s,t \geq 2$. The Fano plane is the unique generalized triangle of
order two. There exists a unique (self-dual) generalized
quadrangle of order two; this object that we already spoke about
in the introduction plays a crucial role in the description of the
generalized Pauli group of two qubits \cite{Metod2,plasan}. It can
be easily checked that the Heawood graph (the incidence graph of
the Fano plane, Fig.\,2) is the generalized hexagon of order
$(1,2)$.
There are also two generalized hexagons of order two,
the so-called split Cayley hexagon and its dual
\cite{Polster1}--\cite{Maldeghem}, \cite{Tits}. Our main concern
here is the former one, also known as the $G_{2}(2)$ hexagon
because the Chevalley group $G_{2}(2)$ is basically its
automorphism group and in what follows simply referred to as the
Hexagon. It has 63 points and 63 lines; every point is contained
in three lines and every line contains three points. One of its
stunning pictures created by the method of ``finite pottery"
\cite{Polster1,Schroth} is reproduced in Fig.\,4.

\begin{figure}[pth!]
\centerline{\includegraphics[width=14.8truecm,clip=]{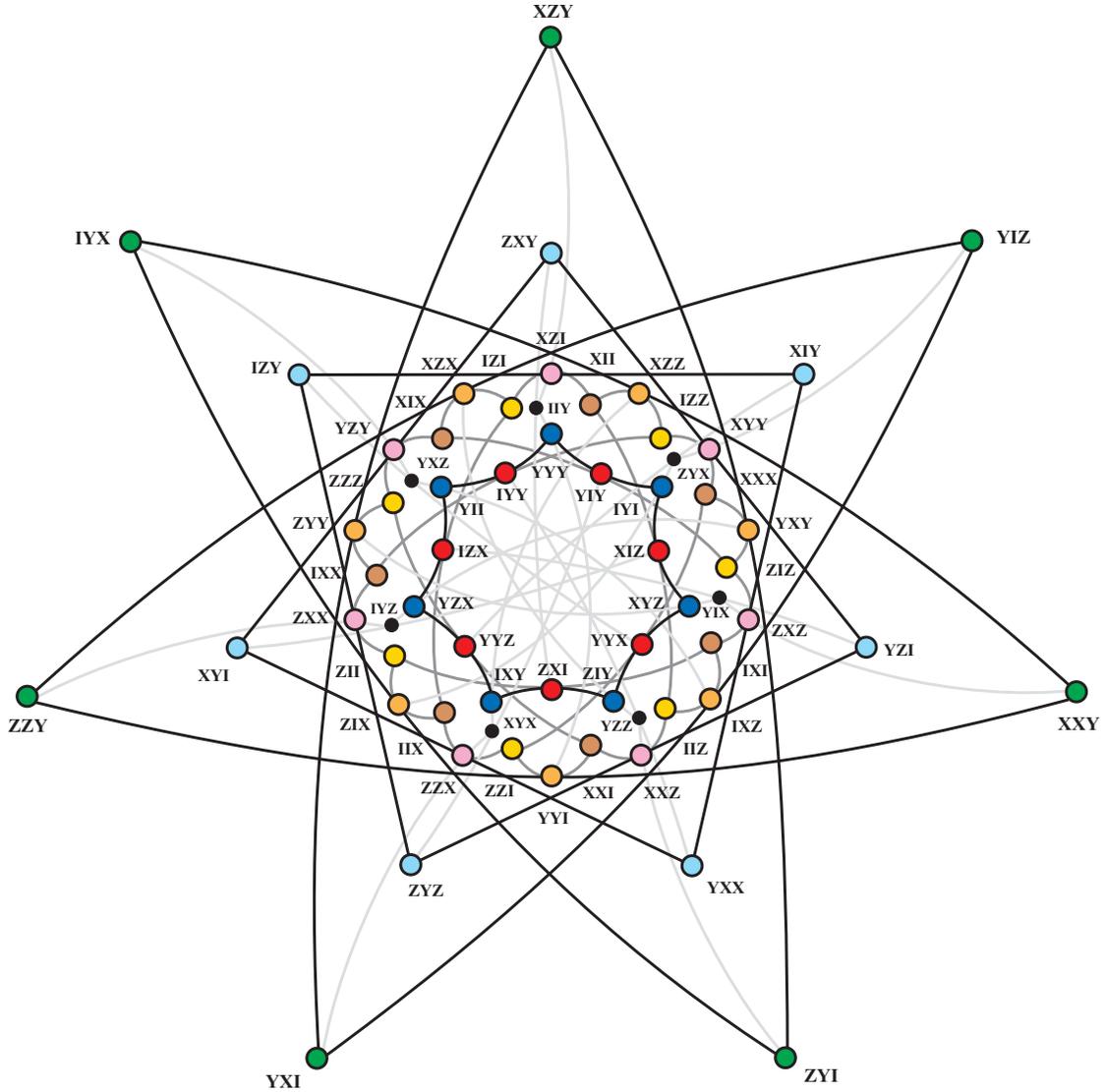}}
\vspace*{0.5cm} \caption{A diagrammatic illustration of the
structure of the split Cayley hexagon of order two (after
\cite{Polster1}--\cite{PSM}) and how this geometry grasps the core
features of the algebra of the real three-qubit Pauli matrices by bijectively associating the latter
with the points of the hexagon.
The points themselves are represented by circles whose
interior is colored in nine different ways reflecting the nine
orbits of an automorphism of order seven; the lines are drawn as black, light and dark grey
curves/joints. The remaining symbols and notation are explained
in the text.}
\end{figure}

Since the Pauli group on three qubits contains $63$ nontrivial
real operators and the Hexagon is endowed with $63$ points/lines
as well, one is immediately tempted to conjecture that the two
objects have something to do with each other. This conjecture is
indeed fairly justified by the facts that the geometry behind the
commutation algebra of the {\it complex} three-qubit Pauli group
is that of the so-called symplectic polar space of order two and
rank three \cite{Metod1,hav} and that our Hexagon lives fully
embedded in this space \cite{thasmal}. Given the seven-fold
symmetry of the Hexagon under which its $63$ points are organized
precisely into $9$ orbits with $7$ elements each (see Fig.\,4) and
the fact that we have an automorphism ${\cal D}(\alpha)$ of order
seven acting on $63$ operators of the Pauli group which gives rise
to the same orbit structure, all we have to do is to find the
algebraic counterpart of the basic incidence relations of the
Hexagon and verify the remaining structural relations.

In order to carry out this program, we first notice that the
Hexagon contains a copy of the Heawood graph (the
sub-configuration in Fig.\,4 consisting of the yellow and brown
circles and the corresponding dark grey lines --- compare with
Fig.\,2). According to our labelling scheme in terms of the
ordered sets ${\cal P}$ and ${\cal L}$, we notice that for the set
$\{$$ZZI$, $IIX$, $ZII$, $IXX$, $ZZZ$, $XIX$, $IZI$, $XII$, $IZZ$,
$XXX$, $ZIZ$, $IXI$, $IIZ$, $XXI$$\}$ of consecutive points of this
graph (Fig.\,2) the operators corresponding to neighboring points
commute. Moreover, according to Table 1, after multiplying these
commuting operators we get symmetric combinations (represented in
Fig.\,4 by the pink and orange circles) which again commute with
their parents. In this way we obtain 28 points of symmetric
combinations grouped into $14$ lines, each containing three
pairwise commuting operators such that the product of any two of
them yields (up to a sign) the third one. It is therefore
reasonable to regard this property as the basic one of incidence
governing the structure of our representation of the Hexagon. One
further sees that the $14$ lines we have just described are of the
point/line/flag type (e.\,g., the triple $\{$$XXI$,$ZZI$,$YYI$$\}$).
These lines are called H1-lines \cite{PSM}, or lines of Coxeter
type \cite{thasmal}. We have altogether 21 such lines (represented
in Fig.\,4 by dark grey arcs), with the remaining $7$ of them
connecting points related to commuting operators of point-line
type to the remaining $7$ symmetric operators (flags); these
latter operators are represented by red circles in Fig.\,4. Having
employed the basic relations of incidence ({\it aka} up-to-a-sign
multiplications producing triples of pairwise commuting
operators), we have thus managed to establish the correspondence
between $35$ points of the Hexagon and  $35$ symmetric
combinations on the one hand, and between $21$ H1-lines and $21$
triples of symmetric combinations on the other hand. As a
consistency check, one readily sees that the order seven
automorphism ${\cal D}(\alpha)$ rotates the operators
corresponding to the yellow, brown, pink, orange and red points in
the clockwise direction.

We still have to take care of the remaining $28$ points
(anti-flags) of the Hexagon that answer to antisymmetric
combinations of operators, as well as for the remaining $42$
H2-lines \cite{PSM}, also known as lines of Heawood type
\cite{thasmal}.\footnote{The names of the two types of lines are
dictated by the fact that after removing the points of flag type
from the Hexagon we are left with two connected graphs --- the
Heawood graph and the Coxeter graph (see below); the edges of the
former/latter corresponding to the lines of Heawood/Coxeter type.}
Let us deal first with the lines. These lines  are of the flag/anti-flag/anti-flag
type and their explicit construction is as
follows. Take a (Fano) line $L$ and a (Fano) point $p$ lying on
it. This gives rise to the flag $\{p,L\}$. Consider now the other
two points $q$ and $r$ lying on the line $L$, and the other two
lines $M,N$ passing through the point $p$. Clearly, the two pairs
$\{q,M\},\{r,N\}$ and $\{q,N\},\{r,M\}$ are two pairs of anti-flags.
Hence, the flag $\{p,L\}$ gives rise to {\it two} H2-lines of the
form $\{\{p,L\},\{q,M\},\{r,N\}\}$ and $\{\{p,L\},\{q,N\},\{r,M\}\}$. Since we
have $21$ flags the number of such lines is $42$. In our
representation in terms of operators of the Pauli group, these
lines are represented by pairwise commuting triples containing two
antisymmetric operators and a symmetric one with the property that
any pair from the triple produces the remaining one under an
up-to-a-sign multiplication. Given the representation of the Fano
plane (Fig.\,1, left, and Fig.\,2) with its points labelled by the
elements of the set ${\cal P}$ and its lines by those from ${\cal
L}$,  it is a rather straightforward task to obtain all of these
$42$ triples, and to check our basic relations of incidence. As an
illustrative example, one takes the point $p$ represented by the
operator $ZII$ and the line $L$ represented by the one $IXX$ (see
Fig.\,1, left). From  Table 1 one reads that they define the flag
$\{p,L\}$ represented by the operator $ZXX$. The remaining two
points on $L$ are $q$ and $r$ represented by $IZZ$ and $ZZZ$,
respectively. The lines $M$ and $N$ going through $p$ correspond,
respectively, to the operators $IIX$ and $IXI$. Hence, the
operator representatives of the two pairs of anti-flags
$\{q,M\},\{r,N\}$ and $\{q,N\},\{r,M\}$ are $IZY,ZYZ$ and $IYZ,ZZY$,
respectively. In this way we have obtained two H2-lines
represented by the triples $\{ZXX$, $IZY$, $ZYZ\}$, and $\{ZXX$,
$IYZ$, $ZZY\}$. The elements of these triples are pairwise
commuting and the product of any two from a given triple indeed
yields, up to a sign, the remaining one. Finally, we observe that
exactly one half of these lines (represented by black
arcs/segments in Fig.\,4) enjoy the property of connecting two
distinct points/operators from the same orbit.

When it
comes to the remaining points of the Hexagon, all what we need is
to make use of the special footing of the seven element set of
antisymmetric matrices given by Eq.\,(\ref{shift6}). For if we
look at Fig.\,4 carefully, the points (related to Fano anti-flags)
represented by black bullets are also found to be special because
the H2-lines passing through them (represented by light grey
curves/arcs in Fig.\,4) encompass not only all $21$ points related
to flags (red, pink and orange circles) but also  all the
remaining $21$ points related to anti-flags (dark blue, light blue
and green circles). None of the remaining points related to
anti-flags exhibit this property. It is, therefore, natural to
associate the points represented by black bullets with the
matrices from ${\cal L}^{\prime}$ (Eq.\,(\ref{shift6})). Given
this fact, and taking into account the structure of H2-lines, the
seven-fold symmetry of the Hexagon and our up-to-a-sign
multiplication recipe, after some experimenting we finally arrive
at the full bijection between our real generalized three-qubit
Pauli matrices and the points of the Hexagon as explicitly shown
in Fig.\,4.

To gain a fuller appreciation of the above-described geometrical
picture of our operator algebra, we have to mention a very
important representation of the Hexagon as a sub-geometry of
$PG(5, 2)$, the five-dimensional projective space over the Galois
field of two elements \cite{Polster1,mald,pick}. In this
representation 63 points of the Hexagon coincide with 63 points of
$PG(5, 2)$, 63 lines of the Hexagon form a special subset out of
the total of 651 lines of  $PG(5, 2)$, and the 35 points of the
Hexagon associated with the {\it symmetric} combinations of the
operators are nothing but the points of a {\it Klein quadric} in
$PG(5, 2)$. And as there exits a bijection between this special
set of points of $PG(5, 2)$ and the 35 lines of $PG(3,2)$, which
completely encodes this projective three-space (see, e.\,g.,
\cite{hirsch}), properties of our symmetric combinations are thus
seen to be also embodied in the configuration of the lines of this
smallest projective space.

We shall conclude this section with the following important
observation. Let us pick up one point represented by a black
bullet. This point is associated with a unique set of seven points
(including the point itself) which are located on the three lines
passing through the point in question. In terms of our bijection,
we thus have seven (one per each black bullet) distinguished sets
of operators, viz.
\beq {\cal
B}_1=\{YZZ,XXZ,ZYI,YXX,IYY,ZIY,XZX\}, \eeq \beq {\cal
B}_2=\{YIX,ZXZ,XXY,YZI,IZX,XYZ,ZYY\}, \eeq \beq {\cal
B}_3=\{ZYX,XYY,YIZ,XIY,YYZ,IYI,ZIX\}, \eeq \beq {\cal
B}_4=\{IIY,XZI,XZY,ZXY,ZXI,YYY,YYI\}, \eeq \beq {\cal
B}_5=\{YXZ,YZY,IYX,IZY,YYX,YII,IXZ\}, \eeq \beq {\cal
B}_6=\{IYZ,ZXX,ZZY,XYI,XIZ,YZX,YXY\}, \eeq \beq {\cal
B}_7=\{XYX,ZZX,YXI,ZYZ,YIY,IXY,XZZ\}. \eeq
Let us add to this list
the two sets ${\cal P}$ and ${\cal L}$, renamed as \beq {\cal
B}_8={\cal P},\qquad {\cal B}_9={\cal L}. \eeq \noindent After
lengthy, yet straightforward, calculations one can verify that
each of them consists of pairwise commuting operators and when
taken together they give rise to a maximum ($d+1 = 8+1 = 9$) set
of mutually unbiased bases (MUBs) in this particular three-qubit
Hilbert space.

\section{The subgroup $PSL_{2}(7)$ of the full automorphism group of the Hexagon}
The Hexagon has a high degree of symmetry. By a symmetry we mean a
bijection sending points to points such that this transformation
induces a bijection between the line sets as well. We have already
encountered the seven-fold symmetry of the Hexagon related to the
permutation ${\alpha}$ and its $8\times 8$ matrix representation
${\cal D}(\alpha)$ of Eq.\,(10). It is known that the full symmetry
group of the Hexagon is of order $12096$, being basically
isomorphic to the Chevalley group $G_2(2) $
\cite{Schroth,Polster2}. In this paper, however, we are merely
interested in an important subgroup of the full automorphism group
of the Hexagon realized on three qubits. This subgroup, which is
directly related to the one of the Fano plane, is Klein's group
$PSL_2(7)$ of order $168$. In the following section, a $8\times
8$ representation of this subgroup on three qubits
will play an important role in illustrating how the geometry
of the Hexagon manifests itself in the physics of certain stringy
black hole solutions.

The presentation of $PSL_2(7)$ we have found convenient \cite{Conway} is that based on three generators $\alpha,\beta$ and $\gamma$
\beq
PSL_2(7)\equiv\{\alpha,\beta,\gamma ~\vert~ {\alpha}^7={\beta}^3={\gamma}^2=
{\alpha}^{-2}{\beta}{\alpha}{\beta}^{-1}=(\gamma\beta)^2=(\gamma\alpha)^3=1\}.
\label{psl27}
\eeq
In the following two possible ways of realizing $PSL_2(7)$ in
terms of permutations of the symbols $1,2,\dots, 7$  will be
needed. They are \beq \alpha\equiv (1234567),\qquad
\beta\equiv(124)(365)(7),\qquad \gamma\equiv (12)(36)(4)(5)(7),
\label{1} \eeq \noindent and \beq {\alpha}^{\prime}\equiv
(1234567),\qquad{\beta}^{\prime}\equiv
(1)(235)(476),\qquad{\gamma}^{\prime}\equiv (1)(3)(4)(25)(67).
\label{2} \eeq \noindent Our convention for multiplying
permutations is from right to left (i.\,e., opposite to that
adopted in Conway and Sloane \cite{Conway}). One can check that
the generators $\alpha,\beta,\gamma$ and
${\alpha}^{\prime},{\beta}^{\prime},{\gamma}^{\prime}$ satisfy the
defining relations of Eq.\,(\ref{psl27}). An important observation
is that if we permute the points of the Fano plane by $PSL_2(7)$
transformations generated by the set $\alpha,\beta,\gamma$ then
this implies a permutation on the lines generated  by the elements
${\alpha}^{\prime},{\beta}^{\prime},{\gamma}^{\prime}$. Hence, the
set defined by Eq.\,(\ref{1}) generates transformations for the
points of the Fano plane and the set of Eq.\,(\ref{2}) for the
points of its dual (see Fig.\,1).

What we need is  an $8\times 8$ representation of $PSL_2(7)$
acting on three-qubits.
It is clear that the $8\times 8$ dimensional representation we are looking for is the one for which
$\alpha$ is represented by ${\cal D}(\alpha)$ of Eq.\,(10).
For the remaining generators, we have found the following representatives
\beq
{\cal D}(\beta)=C_{12}C_{21}=\begin{pmatrix}I&0&0&0\\0&0&0&I\\0&I&0&0\\0&0&I&0\end{pmatrix},
\label{Dbeta}
\eeq
\noindent
\beq
{\cal D}(\gamma)=C_{21}(I\otimes I\otimes Z)=\begin{pmatrix}Z&0&0&0\\0&0&0&Z\\0&0&Z&0\\0&Z&0&0\end{pmatrix},
\label{Dgamma}
\eeq
with the CNOT operations defined by Eq.\,(19).
It is straightforward to check that ${\cal D}(\alpha), {\cal D}(\beta)$ and ${\cal D}(\gamma)$ all satisfy the defining relations of Eq.\,({\ref{psl27}).
It is important to realize that in this way we managed to realize Klein's group
$PSL_2(7)$ on three qubits in terms of CNOT operations and a single ``phase gate" $Z$. (The ``phase" change in this case is just a sign change in the third qubit.)
Hence, Klein's group operates quite naturally on the set ${\cal M}\equiv\{\pm I,\pm X,\pm Y,\pm Z\}$ of the real operators of the Pauli group.

Since the order seven automorphism ${\cal D}(\alpha)$ is by its
very construction a symmetry, in order to demonstrate the
$PSL_2(7)$ symmetry of the Hexagon we have to check the action of
the remaining generators ${\cal D}(\beta)$ and ${\cal D}(\gamma)$
on the three-qubit operators associated with the points of the
Hexagon. To this end, it is convenient to replace our original
labelling scheme (which was motivated by and based on the Polster
and his co-workers' labelling of the points and lines of the
Hexagon \cite{Polster1}--\cite{PSM}) by the one shown in Table 2.
\begin{table}[h]
\begin{center}
\caption{The matrix products between the generalized Pauli operators from the two distinguished sets in terms of the new labelling scheme (compare with Table 1).}
\vspace*{0.5cm}
\begin{tabular}{||l|ccccccc||}
\hline \hline
&&&&&&&\\[-.3cm]
L/P & $h_1$ & $h_2$ & $h_3$ & $h_4$ & $h_5$ & $h_6$ & $h_7$ \\[.1mm]
\hline
&&&&&&&\\[-.3cm]
$i_1$ & $a_1$&
$b_1$
& $c_1$
& $d_1$  &  $e_1$  & $f_1$ & $g_1$ \\
$i_2$ & $g_2$ & $a_2$ & $b_2$ & $c_2$ & $d_2$ & $e_2$ & $f_2$ \\
$i_3$ & $f_3$ & $g_3$ & $a_3$ & $b_3$ & $c_3$ & $d_3$ & $e_3$ \\
$i_4$ & $e_4$ & $f_4$ & $g_4$ & $a_4$ & $b_4$  &  $c_4$  &  $d_4$ \\
$i_5$ & $d_5$ & $e_5$ & $f_5$ & $g_5$ & $a_5$ & $b_5$ & $c_5$ \\
$i_6$ & $c_6$ & $d_6$ & $e_6$ & $f_6$ & $g_6$ & $a_6$ & $b_6$ \\
$i_7$ & $b_7$ & $c_7$ & $d_7$ & $e_7$ & $f_7$ & $g_7$ & $a_7$ \\
\hline \hline
\end{tabular}
\end{center}
\end{table}
So, for example, the set $\{g_k ~\vert ~ k = 1, 2,\dots,
7\}$ corresponds to our set ${\cal L}^{\prime}$ of operators of
Eq.\,(17) playing a special role (black bullets in Fig.\,4) and
the lines of the Hexagon can be written in a nice compact notation
as follows \beq A_k=\{g_{k+1},c_{k+3},a_{k+4}\},
B_k=\{f_{k-2},b_k,g_{k+1}\}, D_k=\{d_{k-1},g_{k+1},e_{k+2}\},
\label{ABD} \eeq \noindent \beq C_k=\{c_{k-2},b_k, c_{k+1}\},
E_k=\{e_{k+1},e_{k+3},a_{k+4}\}, F_k=\{d_{k-1},f_{k+1},f_{k+2}\},
\label{CFE} \eeq \noindent \beq G_k=\{d_{k-1},h_{k+2}, i_{k-1}\},
H_k=\{a_{k+4},h_{k+4},i_{k+4}\}, I_k=\{b_k,h_{k+1},i_k\}.
\label{GHI} \eeq \noindent

Just to familiarize ourselves with this new notation, we pick up one line of the Hexagon, say $C_1=\{c_6,b_1,c_2\}$.
According to Tables 1--2, this line
is an H2-line of the form $\{ZYI,$ $ZIX,$ $IYX \}$ depicted in Fig.\,4 by the colour combination green-orange-green.
In a similar way one finds that the line $G_3=\{d_2,h_5,i_2\}=\{IYY,IZZ,IXX\}$,
with the colour combination red-yellow-brown, is an H1-line.
Obviously, the $42$ lines of type $A,B,C,D,E,F$  are H2-lines, and the $21$
ones of $G,H,I$ type are H1-lines.
The lines containing the special set ${\cal L}^{\prime}$
are the ones of $A,B,D$;
As already remarked in the previous section,  (operators located on) these lines (represented in Fig.\,4 by light grey acrs) give rise to a maximum set of MUBS.

Let us demonstrate the $PSL_2(7)$ symmetry of the Hexagon.
Since we have already proved its seven-fold symmetry, we only have to check how the generators ${\cal D}(\beta)$ and ${\cal D}(\gamma)$ act on the points and lines
of the Hexagon.
Let us consider first the action of ${\cal D}(\beta)$.
We introduce the notation
\beq
(rst)_{ijk}\equiv(r_is_jt_k)(s_it_jr_k)(t_ir_js_k),\quad
(rst)_i\equiv(r_is_it_i),\quad i,j,k=1,2,\dots, 7,
\label{jel}
\eeq
where $(rst)$, as usual, refers to the cyclic permutation of the symbols $rst$.
More generally, we can consider expressions like
\beq
(rst)(uvw)_{(ijk)(lmn)(p)}\equiv (rst)_{ijk}(rst)_{lmn}(rst)_p
(uvw)_{ijk}(uvw)_{lmn}(uvw)_p
\eeq
\noindent
where $(ijk)(lmn)(p)$ is a permutation of the ordered set $\{1,2,\dots, 7\}$ in a cycle notation.
Terms like $(r)_{ijk}$ have no meaning hence they are not included.
Thus, for example, $(uvw)(r)_{(ijk)(l)}$ stands for $(uvw)_{ijk}(uvw)_lr_l$.
The action of ${\cal D}(\beta)$ on the points and lines of the Hexagon is thus given by
\beq
(adb)(cef)(g)(i)_{(1)(253)(467)},\quad (h)_{(142)(356)(7)},
\label{betapoints}
\eeq
and
\beq
(ADB)(CEF)(GIH)_{(142)(356)(7)},
\label{betalines}
\eeq
respectively.
Notice that this action is tied to the {\it inverse}
of the basic permutations ${\beta}$ and ${\beta}^{\prime}$
of Eqs.\,(\ref{1})--(\ref{2}) related to  the corresponding action on the sets
${\cal P}$ ($\{h_k ~ |~ k=1,2,\dots, 7$\}) and ${\cal L}$ ($\{i_k ~ |~ k=1,2,\dots, 7$\}) labelling the points and lines of the Fano plane.

Let us now look at the action of ${\cal D}(\gamma)$.
A straightforward calculation shows that the points transform in the following way
\beq
(a_1b_1)(d_1)(c_1f_1)(e_1)(g_1),\quad (a_3d_3)(b_3)(c_3)(e_3)(f_3g_3),
\label{13}
\eeq
\beq
(a_2d_5)(d_2a_5)(b_2b_5)(c_2g_5)(g_2e_5)(e_2f_5)(f_2c_5),
\label{ko}
\eeq
\beq
(a_4)(b_4)(d_4)(c_4g_4)(e_4f_4),\quad(a_6d_7)(d_6b_7)(b_6a_7)(c_6c_7)(e_6g_7)(g_6f_7)(f_6e_7),
\label{467}
\eeq
\beq
(h_1h_2)(h_3h_6)(h_4)(h_5)(h_7),\quad (i_1)(i_2i_5)(i_3)(i_4)(i_6i_7).
\label{ih}
\eeq
It is important to stress here that the quantities in
Eqs.\,(\ref{13})--(\ref{ko}), as well as $i_1,i_2,i_3,$ and $i_5$,
transform with a sign change; thus, for example, $(a_1b_1)$ and
$(i_1)$ refer to the transformations $a_1\leftrightarrow -b_1$ and
$i_1\leftrightarrow -i_1$. However, since our labelling of the
points of the Hexagon by the real Pauli operators is only {\it up
to a sign}, this additional action of the group ${\bf Z}_2$ is not
of current interest for us. Notice that the transformation rules
given by Eqs.\,(\ref{13})--(\ref{467}), as well as of those of the
points $\{i_k\}$, can be expressed by the permutations of the form
$(1)(3)(25)(4)(67)$, which is just ${\gamma}^{\prime}$ of
Eq.\,(\ref{2}). Moreover, the transformation property
$(ab)(d)(cf)(e)(g)$ of Eq.\,(\ref{13}) and the one of the points
$\{h_k\}$ is associated with $(12)(4)(36)(5)(7)$, i.\,e., with the
permutation ${\gamma}$ of Eq.\,(\ref{1}). Of course, the
transformations on the sets $\{i_k\}$ and $\{h_k\}$ are again the
ordinary ones of the points and lines of the Fano plane.

Finally, we verify that the action of ${\cal D}(\gamma)$ on the
points also induces a bijection on the lines, which is of the form
\beq (A_1D_3)(A_2F_1)(A_3C_6)(A_4C_1)(A_5F_6)(A_6D_4)(A_7),
\label{egy} \eeq \noindent \beq
(B_1E_4)(B_2B_5)(B_3C_3)(B_4C_4)(B_6E_3)(B_7D_7)(C_2C_5)(C_7F_7),
\label{ketto} \eeq \beq
(D_1E_2)(D_2F_2)(D_5F_5)(D_6E_5)(E_1F_3)(E_6F_4)(E_7),
\label{harom} \eeq \noindent \beq
(G_1H_2)(G_2)(G_3H_1)(G_4H_6)(G_5)(G_6H_5)(G_7I_7), \label{negy}
\eeq \beq (H_3I_6)(H_4I_1)(H_7)(I_2I_5)(I_3)(I_4). \label{ot} \eeq
We see that under both ${\cal D}(\beta)$ and ${\cal D}(\gamma)$
the sets of H1- and H2-lines are left invariant. Hence, the
$PSL_2(7)$ subgroup of the full automorphism group of the Hexagon
preserves the splitting of its lines into these two subclasses.
The same is also true for the two distinguished subsets of points.
This is a very important finding which tells us that this subgroup
also preserves the sub-configuration of the Hexagon formed by its
28 anti-flag points and 42 H2-lines which is isomorphic --- as
depicted in Fig.\,5 --- to the famous Coxeter graph \cite{cox},
the second of the two remarkable graphs living in the Hexagon (see
also \cite{PSM}). We shall make use of this result in the next
section. For the sake of completeness, we
here only add that the Hexagon contains 36 different copies of the
Coxeter graph and the same number of copies of the Heawood graph
as well.

\begin{figure}[t]
\centerline{\includegraphics[width=10truecm,clip=]{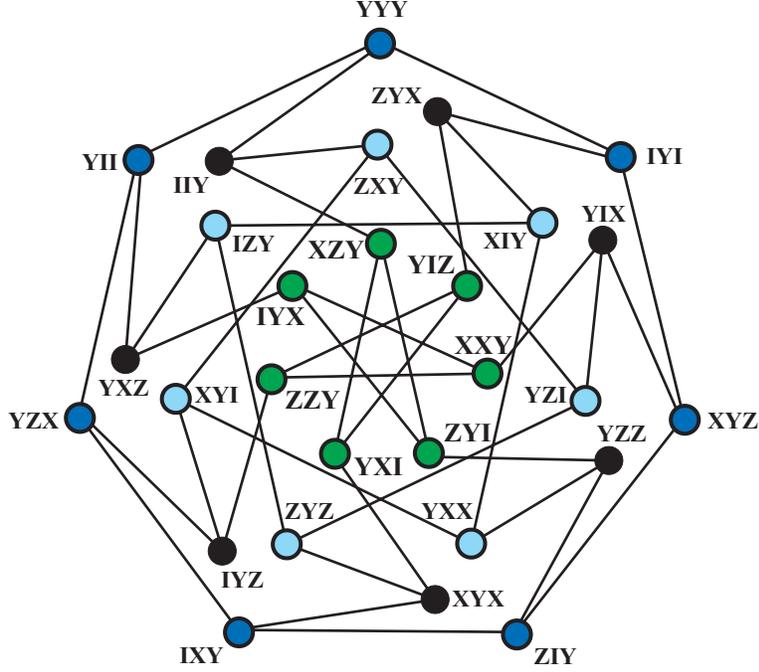}}
\caption{The Coxeter graph, in a form showing its automorphism of
order seven, as a subgraph/subgeometry of the Hexagon (see
Fig.\,4).}
\end{figure}

\section{Coxeter sub-geometry and $E_{7(7)}$-symmetric black hole solutions}
In this section we would like to show how the Coxeter sub-geometry
of the Hexagon manifests itself in the geometry of a certain class
of stringy black hole solutions. In string theory it is well known
that the most general class of black hole solutions in $N=8$
supergravity/M-theory in four dimensions is defined by $56$
charges ($28$ electric and $28$ magnetic), and the entropy formula
for such solutions is related to the square root of the quartic
Cartan-Cremmer-Julia $E_{7(7)}$ invariant
\cite{Kol}--\cite{Cremmer} \beq S=\pi\sqrt{\vert J_4\vert}.
\label{entropy} \eeq \noindent Here, the Cremmer-Julia form of
this invariant \cite{Cremmer} depends on the antisymmetric complex
$8\times 8$ central charge matrix ${\cal Z}$, \beq J_4={\rm
Tr}({\cal Z}\overline{{\cal Z}})^2-\frac{1}{4}({\rm Tr}{\cal
Z}\overline{{\cal Z}})^2+4({\rm Pf}{\cal Z}+{\rm
Pf}\overline{{\cal Z}}), \label{cremmerform} \eeq \noindent where
the overbars refer to complex conjugation. The definition of the
Pfaffian is \beq {\rm Pf}{\cal Z}=\frac{1}{2^4\cdot
4!}{\epsilon}^{ABCDEFGH}{\cal Z}_{AB}{\cal Z}_{CD}{\cal Z}_{EF}{\cal Z}_{GH}. \label{Pfaffian}
\eeq \noindent An alternative form of this invariant is the one
given by  Cartan \cite{Cartan} \beq J_4=-{\rm
Tr}(xy)^2+\frac{1}{4}({\rm Tr}xy)^2-4({\rm Pf}x+{\rm Pf}y).
\label{cartanform} \eeq \noindent Here, the $8\times 8$ matrices
$x$ and $y$ are antisymmetric ones containing  $28$ electric and
$28$ magnetic charges which are integers due to quantization.
These charges are related to some numbers of membranes wrapping
around the extra dimensions these objects live in \cite{Becker}.
The relation between the Cremmer-Julia and Cartan forms has been
established \cite{Larsen} by using the relation \beq {\cal
Z}_{AB}=-\frac{1}{4\sqrt{2}}(x^{IJ}+iy_{IJ})({\Gamma}^{IJ})_{AB},
\label{relation} \eeq \noindent where summation through the
indices $I,J$ is implied. Here
$({\Gamma}^{IJ})_{AB}$ are the generators of the $SO(8)$ algebra,
where $(IJ)$ are the vector indices ($I,J=0,1,\dots, 7$) and
$(AB)$ are the spinor ones ($A,B=0,1,\dots, 7$). Triality of
$SO(8)$ ensures that we can transform between its vector and
spinor representations. A consequence of this is that we can also
invert the relation of Eq.\,(\ref{relation}) and express
$x^{IJ}+iy_{IJ}$ in terms of the matrix of central charge ${\cal
Z}_{AB}$.

Since the matrices $x^{IJ}+iy_{IJ}$ and ${\cal Z}_{AB}$ are both
$8\times 8$ antisymmetric ones with $28$ independent components,
it is natural to expect that the Coxeter sub-geometry of the
Hexagon should provide additional insight into the structure of
the $E_{7(7)}$ symmetric black hole entropy formula. In order to
see that this is indeed the case, all we have to do is to recall
from Sec.\,3 that the elements of our distinguished set ${\cal
L}^{\prime}$ of Eq.\,(17), corresponding to the points of the
Hexagon represented by black bullets (see Fig.\,4), are related to
a seven dimensional Clifford algebra. Moreover, commutators of the
form $[{\Gamma}_j,{\Gamma}_k]$, $j,k=1,2,\dots, 7$, give rise to
the $21$ generators of an $SO(7)$ subalgebra. Hence, the $28$
antisymmetric matrices $({\Gamma}^{IJ})_{AB}$ occurring in
Eq.\,(56) should  be related to the $28$ ones of the Coxeter set
(see Eqs.\,(14)--(17) and Fig.\,5). A natural choice for
relating these two sets is \beq \{g_k\}\leftrightarrow
{\Gamma}^{0k},\qquad [g_j,g_k]\leftrightarrow {\Gamma}^{jk}.
\label{kapcsolat} \eeq \noindent According to our labelling scheme
of Table 2, we see that the $ 28$ $SO(8)$ generators
$({\Gamma}^{IJ})_{AB}$ are divided into the four sets $\{g_k\}$,
$\{b_k\}$, $\{e_k\}$ and $\{f_k\}$, $k=1,2,\dots,7$. An explicit
relation between these quantities is \beq
-{\Gamma}^{IJ}=\begin{pmatrix}0&g_1&g_2&g_3&g_4&g_5&g_6&g_7\\
-g_1&0&e_6&c_4&-f_3&f_7&-c_2&-e_5\\
-g_2&-e_6&0&e_7&c_5&-f_4&f_1&-c_3\\
-g_3&-c_4&-e_7&0&e_1&c_6&-f_5&f_2\\
-g_4&f_3&-c_5&-e_1&0&e_2&c_7&-f_6\\
-g_5&-f_7&f_4&-c_6&-e_2&0&e_3&c_1\\
-g_6&c_2&-f_1&f_5&-c_7&-e_3&0&e_4\\
-g_7&e_5&c_3&-f_2&f_6&-c_1&-e_4&0\end{pmatrix}. \label{gammacox}
\eeq \noindent Here, the first row and column corresponding to the
entries ${\Gamma}^{0k}$ and ${\Gamma}^{k0}$ are related to the
special set $\{g_k\}$. The entries of the remaining $7\times 7$
block are given by the formula $\frac{1}{2}[g_j,g_k]$. Thus, for
example, according to Tables 1 and 2 we have
$-{\Gamma}^{25}=\frac{1}{2}[g_2,g_5]=\frac{1}{2}[ZYX,XYX]=-YII=-f_4$.
(All the entries of this matrix ${\Gamma}^{IJ}$ are $8\times 8$
matrices, hence the spinor indices $AB$ are left implicit.)

An important consequence of this explicit correspondence is that
the $28$ Gaussian integers (regarded as expansion coefficients for
${\cal Z}$) comprising the antisymmetric charge matrix
$x^{IJ}+iy_{IJ}$ can also divided into seven sets. This division
answers to the similar one of the ``basis" vectors
$\{g_k,e_k,c_k,f_k\}$, $k=1,2,\dots, 7$, corresponding to
${\Gamma}^{IJ}$. For historical reasons (to be explained below) we
chose the {\it dual labelling} for the elements of the charge
matrix $x^{IJ}+iy_{IJ}$ \beq x^{IJ}=\begin{pmatrix}
0&-{\bf a}_7&-{\bf b}_7&-{\bf c}_7&-{\bf d}_7&-{\bf e}_7&-{\bf f}_7&-{\bf g}_7\\
{\bf a}_7&0&{\bf f}_1&{\bf d}_4&-{\bf c}_2&{\bf g}_2&-{\bf b}_4&-{\bf e}_1\\
{\bf b}_7&-{\bf f}_1&0&{\bf g}_1&{\bf e}_4&-{\bf d}_2&{\bf a}_2&-{\bf c}_4\\
{\bf c}_7&-{\bf d}_4&-{\bf g}_1&0&{\bf a}_1&{\bf f}_4&-{\bf e}_2&{\bf b}_2\\
{\bf d}_7&{\bf c}_2&-{\bf e}_4&-{\bf a}_1&0&{\bf b}_1&{\bf g}_4&-{\bf f}_2\\
{\bf e}_7&-{\bf g}_2&{\bf d}_2&-{\bf f}_4&-{\bf b}_1&0&{\bf c}_1&{\bf a}_4\\
{\bf f}_7&{\bf b}_4&-{\bf a}_2&{\bf e}_2&-{\bf g}_4&-{\bf c}_1&0&{\bf d}_1\\
{\bf g}_7&{\bf e}_1&{\bf c}_4&-{\bf b}_2&{\bf f}_2&-{\bf a}_4&-{\bf d}_1&0
\end{pmatrix},
\label{x}
\eeq
\noindent
\beq
y_{IJ}=\begin{pmatrix}
0&-{\bf a}_0&-{\bf b}_0&-{\bf c}_0&-{\bf d}_0&-{\bf e}_0&-{\bf f}_0&-{\bf g}_0\\
{\bf a}_0&0&{\bf f}_6&{\bf d}_3&-{\bf c}_5&{\bf g}_5&-{\bf b}_3&-{\bf e}_6\\
{\bf b}_0&-{\bf f}_6&0&{\bf g}_6&{\bf e}_3&-{\bf d}_5&{\bf a}_5&-{\bf c}_3\\
{\bf c}_0&-{\bf d}_3&-{\bf g}_6&0&{\bf a}_6&{\bf f}_3&-{\bf e}_5&{\bf b}_5\\
{\bf d}_0&{\bf c}_5&-{\bf e}_3&-{\bf a}_6&0&{\bf b}_6&{\bf g}_3&-{\bf f}_5\\
{\bf e}_0&-{\bf g}_5&{\bf d}_5&-{\bf f}_3&-{\bf b}_6&0&{\bf c}_6&{\bf a}_3\\
{\bf f}_0&{\bf b}_3&-{\bf a}_5&{\bf e}_5&-{\bf g}_3&-{\bf c}_6&0&{\bf d}_6\\
{\bf g}_0&{\bf e}_6&{\bf c}_3&-{\bf b}_5&{\bf f}_5&-{\bf a}_3&-{\bf d}_6&0
\end{pmatrix}.
\label{y}
\eeq
\noindent
Here, we used boldface letters for the coefficients ${\bf a}_J,\dots,{\bf g}_J, J = 0 ,1,\dots, 7$,
not to be confused with the letters $a_j,\dots, g_j, ~j=1,2,\dots,7$, denoting $8\times 8$ matrices.
This labelling can be summarized as
\beq
\{1,2,3,4,5,6,7\}\leftrightarrow\{{\bf a},{\bf b}, {\bf c},{\bf d},{\bf e},{\bf f},{\bf g}\}
\label{conv1}
\eeq
\beq
\{g,e,,c,f\}\leftrightarrow\{70,16,43,25\}.
\label{conv2}
\eeq
\noindent
The symbols on the left-hand side of Eqs.\,(\ref{conv1})--(\ref{conv2})
label the basis vectors, those on the right hand side the expansion coefficients of the central charge matrix ${\cal Z}$.
In other words, we are converting numbers occurring in the sets $\{g_k,e_k,c_k,f_k\}$, $k=1,\dots,7$, to boldface letters in a lexicographic order
and the letters $g,e,c,f$ to numbers according to the rule of Eq.\,(\ref{conv2}).
This means that, for example, we use the notation for the expansion coefficient $x^{36}+iy_{36}$
attaching to the ``basis vector" $-{\Gamma}^{36}\sim-f_5$ the symbol
$-{\bf e}_2-i{\bf e}_5$; that is, $5$ was converted to ${\bf e}$ and the letter
$f$ was converted to the numbers $2$ and $5$ corresponding to the real and imaginary parts of the charge matrix.

In this way we can associate the $28$ points of the Coxeter
sub-geometry of the Hexagon to $28$ Gaussian integers with the
real and imaginary parts corresponding to electric and magnetic
charges characterizing the particular charge configuration of the
black hole solution. Explicitly, according to Tables 1 and 2, the
sets of points represented by the (black, green, light blue, dark
blue) circles in Fig.\,5 can be associated with the ``basis
vectors" $(g_k,c_k,e_k,f_k), k=1,2,\dots,7$. Moreover, according
to Eqs.\,(\ref{gammacox})--(\ref{y}), with these ``basis vectors"
we can associate the corresponding ``expansion coefficients,"
i.\,e., the charge combinations.

By virtue of the seven-fold symmetry of the Coxeter graph we also
have managed to divide the $56$ charges characterizing the
particular black hole solution into $7$ subsets with $8$ elements
each. It can be shown \cite{Ferrara,Levay4} that each of these $7$
sets can be regarded as $7$  three-qubit states with the $8$
elements being their integer-valued amplitudes. The explicit
dictionary in the form of Eqs.\,(\ref{x}) and (\ref{y}) between
the amplitudes of these three-qubit states and the elements of the
charge matrices $x^{IJ}$ and $y_{IJ}$ were first obtained by Duff
and his co-workers \cite{Duminda}. Here, we recovered this result
by a straightforward way from the geometry of the Hexagon. It was
also shown \cite{Ferrara,Levay4} that the transformation
properties of the $56$ charges related to such three-qubit states
under the fundamental representation of the exceptional group
$E_7$ can be described by introducing an entangled state of seven
qubits related to the geometry of the Fano plane. This seven-qubit
system, however, contains merely tripartite entanglement in the
form of our seven three-qubit systems we have just found.

Notice that our present approach to this unusual type of
entanglement is {\it fundamentally} different from those based on
seven qubits. Here we managed to describe the charge
configurations of black hole solutions by using the algebra of
{\it observables} on {\it three}-qubits. By comparison in these
recent approaches such configurations have been described by
special {\it states} on {\it seven}-qubits forming a $56$
dimensional subspace of the state space of {\it seven qutrits}
\cite{Ferrara}.

In the previous section we have shown that the Coxeter
sub-geometry has a $PSL_{2}(7)$ symmetry. Due to the relationship
found between this geometry and the charge configurations  we
expect that this symmetry should also manifest itself in the
black hole entropy formula. That this should be the case is
already known from the mathematical literature \cite{Manivel}.
However, now we can demonstrate this symmetry quite easily. Recall
our $8\times 8$ matrix representation of $PSL_{2}(7)$ generated by
the matrices ${\cal D}(\alpha),{\cal D}(\beta),{\cal D}(\gamma)$.
It is easy to check that these generators are elements of the
group $SO(8)$. Since under an element $O\in SO(8)$ the Pfaffian
transforms as ${\rm Pf}(O{\cal Z}O^t)=({\rm Det O}){\rm Pf}({\cal
Z})={\rm Pf}({\cal Z})$, from Eq.\,(\ref{cremmerform}) one can
immediately check that $J_4$, and so the black hole entropy
formula, is invariant under $PSL_{2}(7)$. This should not come as a
surprise since each of the terms of Eq.\,(\ref{cremmerform}) has
manifest $SU(8)$ symmetry and all of our generators can be
embedded into this subgroup of $E_{7(7)}$. What is not trivial,
however, is the explicit form of this $PSL_{2}(7)$ action on the
charge matrix $x^{IJ}+iy_{IJ}$ which is simply the one induced by
a similar action of conjugation on the basis vectors
${\Gamma}^{IJ}$. This explicit action on the ``basis vectors"
$\{g_k,c_k,e_k,f_k\}, k=1,\dots, 7$, related to ${\Gamma}^{IJ}$
via Eq.\,(58), is \beq {\cal
D}(\alpha):(c)(e)(f)(g)_{(1234567)}, \label{a} \eeq \beq {\cal
D}(\beta):(cef)(g)_{(1)(253)(457)}, \label{b} \eeq \beq {\cal
D}(\gamma):(c_1f_1)(e_1)(g_1),\quad
(c_2g_5)(g_2e_5)(e_2f_5)(f_2c_5),\quad (c_3)(e_3)(f_3g_3),
\label{gam1} \eeq \beq {\cal D}(\gamma): (c_4g_4)(e_4f_4),\quad
(c_6c_7)(e_6g_7)(g_6f_7)(f_6e_7). \label{gam2} \eeq \noindent
Here, we again note that the quantities of Eq.\,(\ref{gam1}))
transform with a sign change, i.\,e., $(c_1f_1)$ means
$c_1\leftrightarrow -f_1$ (see
Eqs.\,(\ref{betapoints}),(\ref{13})--(\ref{467})). By virtue of
the correspondence as given by
Eqs.\,(\ref{relation}),(\ref{gammacox})--(\ref{y}), it is easy to
obtain the corresponding explicit $PSL_{2}(7)$ action on the charge
matrices $x$ and $y$, or, alternatively, on the seven three-qubit
systems built from seven qubits \cite{Ferrara,Levay4}. This
explicit action of the $PSL_{2}(7)$ symmetry can be regarded as a
subgroup of transformations of the full $U$-duality group $E_7({\bf
Z})$. Notice also that we have managed to represent this duality
subgroup in terms of three-qubit CNOT operations creating and
destroying entanglement and a trivial phase gate $Z$.

As the last application of the Coxeter sub-geometry to stringy
black hole solutions, let us consider the canonical form of the
central charge matrix ${\cal Z}$. As it is well known, by a
transformation of the form ${\cal Z}\mapsto U^{t}{\cal Z}U$, where
$U\in SU(8)$, ${\cal Z}$ can be brought into the form \beq {\cal
Z}_{{\rm canonical}}=\begin{pmatrix}z_1&0&0&0\\0&z_1&0&0\\0&0&z_3&0\\0&0&0&z_4\end{pmatrix}\otimes
Y, \label{kanonikus} \eeq \noindent where all four $z_j$ could be
chosen to have the same phase, or three of the $z_i$ could be
chosen to be real \cite{Becker}. Observing that the antisymmetric
combinations $c_1,e_1,f_1,g_1$ occurring in the first row of Table
1 have the following explicit forms \beq
c_1=\begin{pmatrix}1&0&0&0\\0&-1&0&0\\0&0&-1&0\\0&0&0&1\end{pmatrix}\otimes
Y,\quad
e_1=\begin{pmatrix}1&0&0&0\\0&-1&0&0\\0&0&1&0\\0&0&0&-1\end{pmatrix}\otimes
Y, \label{explicit1} \eeq \noindent \beq
f_1=\begin{pmatrix}1&0&0&0\\0&1&0&0\\0&0&-1&0\\0&0&0&-1\end{pmatrix}\otimes
Y,\quad
g_1=\begin{pmatrix}1&0&0&0\\0&1&0&0\\0&0&1&0\\0&0&0&1\end{pmatrix}\otimes
Y,     \label{explicit2} \eeq \noindent we see that obtaining the
canonical form corresponds to the situation of attaching
charges merely to the points of the Coxeter graph  labelled by $ZZY$,
$IZY$, $ZIY$ and $IIY$ (see Fig.\,5).

Let us first assume that the
numbers $z_i$ are all real. Then, employing Eqs.\,(\ref{relation}),(\ref{gammacox})
and (\ref{x}), we get \beq \sqrt{8}{\cal
Z}=x^{01}g_1+x^{34}e_1+x^{57}c_1+x^{26}f_1=-{\bf a}_7g_1+{\bf
a}_1e_1+{\bf a}_4c_1+{\bf a}_2f_1. \label{kancox} \eeq \noindent
Hence, in this case we have \beq z_1=\frac{1}{\sqrt{8}}(-{\bf a}_7+{\bf
a}_2+{\bf a}_1+{\bf a}_4),\quad z_2=\frac{1}{\sqrt{8}}(-{\bf a}_7+{\bf
a}_2-{\bf a}_1-{\bf a}_4), \label{z12} \eeq \noindent \beq
z_3=\frac{1}{\sqrt{8}}(-{\bf a}_7-{\bf a}_2+{\bf a}_1-{\bf a}_4),\quad
z_4=\frac{1}{\sqrt{8}}(-{\bf a}_7-{\bf a}_2-{\bf a}_1+{\bf a}_4),
\label{z34} \eeq \noindent with the entropy formula \beq
S=\pi\sqrt{\vert -4{\bf a}_1{\bf a}_2{\bf a}_4{\bf a}_7\vert}
\label{ent} \eeq \noindent being just the usual one obtained for
four-charge extremal black holes \cite{Becker}. (We remind that
$J_4$ is positive for BPS and negative for non-BPS charge
combinations.)

In the usual
``tripartite-entanglement-of-seven-qubits" interpretation
\cite{Ferrara,Levay4} of the fundamental representation of $E_7$
we have seven three-qubit states with amplitudes ${\bf
a}_{I},\dots, {\bf g}_I, I=0,1,\dots, 7$. In this picture,
obtaining the canonical form is simply restriction to just one of
such three-qubit states (in this case to the one with amplitudes
${\bf a}_I$). This process is equivalent
to the one with $-{\bf a}_7-i{\bf a}_0$ attached to the black
circle labelled by $IIY$, and ${\bf a}_4+i{\bf a}_3$, ${\bf
a}_2+i{\bf a}_5$ and ${\bf a}_1+i{\bf a}_6$ to the circles
labelled by $ZZY$, $ZIY$ and $IZY$, i.\,e., to the corresponding
green, dark blue and light blue elements of Fig.\,5. As explained
elsewhere \cite{Linde,Ferrara,Levay4}, the black hole entropy
formula in this case is related to Cayley's hyperdeterminant
\cite{Duff1}, i.\,e., to the unique triality and $SL(2)^{\otimes
3}$ invariant entanglement measure for three qubits.
Indeed, a straightforward calculation in this case shows that
$J_4=-D({\bf a})$ where the explicit expression for Cayley's hyperdeterminant $D({\bf a})$ is
\begin{eqnarray}
D({\bf a})&=&({\bf a}_0{\bf a}_7)^2+({\bf a}_1{\bf a}_6)^2+({\bf a}_2{\bf a}_5)^2+({\bf a}_3{\bf a}_4)^2
-2({\bf a}_0{\bf a}_7)[({\bf a}_1{\bf a}_6)+({\bf a}_2{\bf a}_5)+({\bf a}_3{\bf a}_4)]\nonumber\\
&-&2[({\bf a}_1{\bf a}_6)({\bf a}_2{\bf a}_5)+({\bf a}_2{\bf a}_5)({\bf a}_3{\bf a}_4)+({\bf a}_3{\bf a}_4)({\bf a}_1{\bf a}_6)]\nonumber\\
&+&4{\bf a}_0{\bf a}_3{\bf a}_5{\bf a}_6+4{\bf a}_1{\bf a}_2{\bf a}_4{\bf a}_7
\end{eqnarray}
\noindent
giving a generalization of Eq.\,(\ref{ent}) as $S=\pi\sqrt{\vert -D({\bf a})\vert}$.

Obviously the action of ${\cal D}(\alpha)$ on the canonical form of ${\cal Z}$
by conjugation gives rise to the permutation of the charges in the form $({\bf abcdefg})$. Combining this with the $PSL_2(7)$ invariance of $J_4$ shows that
in agreement with the well-known result \cite{Ferrara,Levay4,Manivel}
the expression for $J_4$ in terms of the $56$ charges contains (among others) the sum of
seven copies of Cayley's hyperdeterminant.
These terms describe the seven different possible $N=2$, $D=4$ STU truncations 
of the $N=8$ solutions.
Using the explicit dictionary of Eqs.\,(\ref{gammacox})--(\ref{y}) we can alternatively label the points of Fig.\,5 by the $28$ Gaussian integers.
It is easy to check that the charge configurations belonging to a particular truncation of this type correspond to seven sets of four points with mutual distance {\it four}.

After we have clarified the geometrical meaning of the canonical form for ${\cal Z}$, let us try to understand its symmetry properties.
As already explained, we have only eight charges corresponding to the eight
integers ${\bf a}_I, I=0,1,\dots, 7$.
The action of the $PSL_{2}(7)$ subgroup of the duality group on these charges
can be understood from the corresponding action on the
operators $c_1,e_1,f_1,g_1$ (see
Eqs.\,(\ref{conv1})--(\ref{conv2})).
From Eqs.\,(\ref{b}) and (\ref{gam1}) we find that ${\cal D}(\beta)$ and ${\cal D}(\gamma)$,
acting as $(c_1e_1f_1)(g_1)$ and $(c_1f_1)(e_1)(g_1)$,
generate the dihedral group $D_3$, which is isomorphic to the permutation group $S_3$.
The action of $S_3$ on the eight charges in the permutation notation reads
\beq
{\cal D}(\beta): ({\bf a}_0)({\bf a}_7)({\bf a}_3{\bf a}_6{\bf a}_5)({\bf a}_1{\bf a}_2{\bf a}_4),
\label{tr1}
\eeq
and
\beq
{\cal D}(\gamma): ({\bf a}_0)({\bf a}_7)({\bf a}_1)({\bf a}_6)({\bf a}_2{\bf a}_4)({\bf a}_3{\bf a}_5).
\label{tr2}
\eeq
\noindent
At this spoint, we get in touch with the usual interpretation based on seven three-qubit systems \cite{Ferrara,Levay4}. Let us associate with the $8$ charges ${\bf a}_I, I=0,1\dots, 7$,
the three-qubit state
\beq
\vert\psi\rangle\equiv {\bf a}_{000}\vert 000\rangle +{\bf a}_{001}\vert 001\rangle + \dots + {\bf a}_{111}\vert 111\rangle,
\label{state}
\eeq
\noindent
i.\,e., we reinterpreted the charges ${\bf a}_I$ (decimal labeling)
as integer-valued amplitudes ${\bf a}_{abc}$, $a,b,c$ $\in \{0,1\}$ (binary labeling) of a three-qubit state.
Then the transformation rules of Eqs.\,(\ref{tr1}) and ({\ref{tr2}) for the amplitudes correspond to the permutations
$(321)$ and $(12)$ of subsystems (we are labeling qubits from left to right).
Since these permutations
generate the full permutation group on three letters, the corresponding truncation of $J_4$ (i.\,e., Cayley's hyperdeterminant) should be a permutation invariant,
and this is indeed the case. Hence, as another important consistency check, we reproduced the permutation invariance of Cayley's hyperdeterminant from the $PSL_{2}(7)$ symmetry
of Cartan's quartic invariant.

\section{Conclusion}
We have performed a comprehensive analysis of the properties of
the {\it real} generalized Pauli operators of three-qubits to the
extent comparable with the treatment of the (complex) two-qubit
case \cite{Metod2,plasan}. The algebra of the operators is found
to be completely describable in terms of the structure of the Fano
plane and uniquely extendible to the picture of the split Cayley
hexagon of order two. The 63 Pauli operators are bijectively
identified with 63 points of the Hexagon whose 63 lines carry each
triples of operators; these triples are such that the product of
any two of them gives, up to a sign, the third one. A deeper
insight into the structure of the Hexagon is acquired in terms of
one of its automorphism of order 7 that features 9 orbits of 7
points/operators each. The factorization of the set of operators
into 35 symmetric and 28 antisymmetric members finds its natural
explanation in terms of two distinct kinds of points of the
Hexagon: 7 Fano points, 7 Fano lines and 21 Fano flags on the one
hand versus 28 Fano anti-flags on the other hand, respectively.
Triples of collinear operators are of two distinct kinds as well,
according as they lie on the lines of point/line/flag (Heawood)
type or on those of flag/anti-flag/anti-flag (Coxeter) type. The
geometry behind 28 antisymmetric matrices is governed by the
properties of the Coxeter graph and its symmetries. The geometry
behind 21 out of 35 symmetric guys rests on the Heawood graph,
while that of the full set is mimicked by a Klein quadric in a
projective 5-space over $GF(2)$.

Employing this novel finite geometrical language, an alternative
quantum informational view of the structure of the $E_7$-symmetric
black hole entropy formula and a subgroup of its duality symmetry
has been arrived at. This view is based on the {\it algebra of
observables on three-qubits}. This is to be contrasted with the
other ones \cite{Ferrara,Levay4} based on {\it an entangled state
of seven qubits}. In both pictures, however, three-qubits play a
distinguished role. We managed to establish a dictionary between
these complementary approaches. However, there are still many
questions left to be answered. The most exciting one is the
possibility of relating the black hole entropy formula to the {\it
full geometry of the Hexagon}. Indeed, we only managed to relate
black hole solutions to the Coxeter sub-geometry with an explicit
$PSL_{2}(7)$ symmetry. We conjecture that the full $G_2(2)$ symmetry
of the Hexagon should play a role in this black hole context. One
solid piece of evidence in favour of this conjecture is the fact
that since $E_7$ is of rank seven we have $133-7=2\times 63=126$
step operators. According to an important result of M. Koca and
his co-workers \cite{Koca}, the adjoint Chevalley group $G_2(2)$
acts naturally on the (octonionic) root system of $E_7$. Using
this result and our explicit dictionary, it would be really
spectacular to establish a full correspondence between the
geometry of the Hexagon and the duality symmetries of stringy
black hole solutions.

As already remarked, there exist two distinct generalized hexagons
of order two: our Hexagon and its dual. One is naturally tempted
to ask why three-qubits favour the former, not the latter. Since
the algebra of the corresponding operators is encoded in the
structure of the Fano plane, the answer seems to be rather easy:
our Hexagon, as we have seen, contains a sub-hexagon of order ($1,
2$) which is isomorphic to the incidence graph of the Fano plane,
but its twin does not. Yet, there exists another remarkable
structure within our Hexagon which is absent in its counterpart, a
{\it distance-2-ovoid} \cite{Polster1}--\cite{Maldeghem}. A
distance-2-ovoid is a set of non-collinear points such that every
line is incident with exactly one point of that set and for a
generalized hexagon of order $(s,t)$ its cardinality amounts to
$s^{2}t^{2} + st + 1$. In our Hexagon, this is the {\it unique}
configuration of 21 points of flag type (orange, red and pink
circles in Fig.\,4). Why we should be bothered with this set?
Because generalized hexagons and their sub-configurations have
recently been found to be an important source of a variety of {\it
codes} \cite{wismal}, with those arising from distance-2-ovoids
being of specific two-weight type. This link with coding theory
is something which certainly deserves a serious perusal of its own.

As a final note, we would like to draw the reader's attention to
the following puzzling analogy. We have already mentioned in the introduction that
multi-qubit states of smallest orders can geometrically be
modelled by Hopf fibrations. Namely, the $k$-qubit
states, $k = 1, 2, 3$, are intimately connected with the Hopf
sphere fibrations of the

$S^{2^{(k+1)}-1} \stackrel{S^{2^{k}-1}} \longrightarrow S^{2^{k}}$

\noindent type, respectively \cite{Mosseri,Bernevig}. A very
similar situation occurs in the setting provided by thick
generalized $n$-gons.
For if we associate the geometry of three
non-trivial matrices given by
Eq.\,(3) with that of the
generalized digon ($n = 2$) of order two, and take into account
that the two-qubit case is underlaid by the geometry of the
generalized quadrangle ($n = 4$) of order two
\cite{Metod2,plasan}, we also arrive at such a three-fold
correspondence: $n = 2k$, $k = 1, 2, 3$.
Moreover, in both the
cases, remarkably, the sequence ends at $k = 3$; in the former
case due to the fact that there are no natural Hopf fibrations of
higher order, whilst in the latter one because, by the celebrated
result of Feit and Higman \cite{fh}, of non-existence of thick
generalized $n$-gons with $n > 6$ and meeting the constraint $s = t$.
We thus observe a fascinating phenomenon when two substantially different pieces of
mathematics convey the same message: a special footing played by single-qubit, two-qubit and
three-qubit systems within an infinite family of multiple-qubit states.
Is this analogy a mere coincidence, or is there a deep physical reason behind?

\section*{Acknowledgements}
One of us (P.L.) would like to express his gratitude to professor Michael Duff  for the warm hospitality during his recent stay at the Imperial College in London.
His special thanks also go to Leron Borsten and Duminda Dahanayake for useful correspondence.
This work was partially supported by the VEGA grant agency projects Nos. 6070
and 7012. We are extremely grateful to our friend Petr Pracna (Prague) for creating the
electronic versions of all the figures.

\end{document}